**Modern cities emerge as 'super-cells' where enclosed industrial systems are hotspots of goods and services**

Jie Chang[1*], Ying Ge[1], Zhaoping Wu[1], Yuanyuan Du[1], Kaixuan Pan[1], Guofu Yang[1], Yuan Ren[1], Mikko P. Heino[2], Feng Mao[3], Zelong Qu[1], Xing Fan[1], Yong Min[4], Changhui Peng[5], Laura A. Meyerson[6*]

*1 College of Life Sciences, Zhejiang University, Hangzhou 310058, China*

*2 Department of Biological Sciences, University of Bergen, Bergen N-5020, Norway*

*3 School of Geography, Earth and Environmental Sciences University of Birmingham, Birmingham B15 2TT, United Kingdom*

*4 College of Computer Science and Technology, Zhejiang University of Technology, Hangzhou 310014, China.*

*5 Center of CEF/ESCER, Department of Biological Sciences, University of Quebec at Montreal, Montreal H3C 3P8, Canada*

*6 Natural Resources Science, University of Rhode Island, Kingston, RI 02881 USA*

*Authors for correspondence. E-mail: jchang@zju.edu.cn or lameyerson@uri.edu

**SUMMARY**

Prevailing hypotheses recognize cities as 'super-organisms' which both provides organizing principles for cities and fills the scalar gap in the hierarchical living system between ecosystems and the entire planet. However, most analogies between the traits of organisms and cities are inappropriate making the super-organism model impractical as a means to acquire new knowledge. Using a cluster analysis of 15 traits of cities and other living systems, we found that modern cities are more similar to eukaryotic cells than to multicellular organisms. Enclosed industrial systems, such as factories and greenhouses, dominate modern cities and are analogous to organelles as hotspots that provide high-flux goods and services. Therefore, we propose a 'super-cell city model' as more appropriate than the super-organism model. In addition to the theoretical significance, our model also recognizes enclosed industrial systems as functional components that improve the vitality and sustainability of cities.

**INTRODUCTION**

The most significant feature of modern cities (urban-rural complexes) is the emergence of a great number of industrialized functional components. For example, factories, greenhouses, dairy farms and other infrastructure provide goods such as industrial products, food, water, fibre, and services such as environmental pollution control through waste treatment plants[1,2]. Increasingly, functional components are becoming more complex and smarter[3] and are making modern cities distinct from pre-industrial traditional cities in structure, processes, and functions. Pre-industrial traditional cities had only non-industrial components such as restaurants, hotels and workshops (shoemakers, smithies and so on), which were small in size and simple in structure. Although modern cities have brought many benefits to humans, many problems such as environmental pollution and land degradation have also arisen[1,4]. In the Anthropocene, modern cities should enhance both vitality and sustainability simultaneously which means that cities require solutions to provide sufficient goods and services to people while also addressing environmental problems[5] Yet, most approaches are based on 'trial-and-error' methods despite the many advances in the science, planning, engineering, and management of cities [2,6]. Clearly, the theoretical basis for the future of science and engineering in modern cities needs to be strengthened.

Some researchers have recognized that cities share characteristics similar to living systems and have proposed a 'super-organism' hypothesis[6]. In this model, wetlands are analogized as the city's kidneys, green spaces as the city's lungs, and streets as its blood vessels[4]. This model includes cities as a level on the hierarchy of living systems and informs studies such as industrial ecology and urban metabolism[7]. For example, energy consumption in response to urban mass follows the power-law function $Y = aX^\beta$, with the allometric scaling exponent $\beta \sim 5/6$ [8]. The sublinear response as such means that larger cities have a relatively lower energy consumption than smaller cities. This is similar to the sublinear scaling ($\beta \sim 3/4$) of energy metabolism of multicellular organisms (especially animals) to their biomass [6]. The difference (5/6 - 3/4) should be the optimal goal for city development. However, the super-organism model is an imperfect analogy because comparisons of organisms and cities are not appropriate for most features. In terms of morphology and structure, for example, an organism has only a few organs and the quantity of organs is often fixed while a city can have many functional components of the same kind. Another structural difference is that it is important to optimize the spatial distribution of functional components in cities to improve the provision goods and services but the position of organs cannot change in organisms. In terms of vitality, the walking speed of humans in response to city size is superlinear ($\beta > 1$), while the heart rate of mammals in response to body size is sublinear[9]. The discrepancies described above suggest a lack of common principles between cities and organisms. Furthermore, engineers and managers argue that such analogies with organisms are often inappropriate and provide limited insights for urban planning and management[10,11]. This suggests that to build a more credible and useful model, cities need to be analogized to a more appropriate type of living system.

    Here, we first analyse the typical characteristics of modern cities and industrial systems. All industrial systems are enclosed and covered by manmade films such as glass, plastic, cements, and other materials, including enclosed ecosystems and enclosed non-ecosystems. We then report the relationships between modern cities and several levels

of living systems, including organisms and organs and eukaryotic cells and their organelles, in the hierarchical living system. We identified 15 traits (7 quantitative and 8 qualitative) that cover the morphology, structure, processes, and functions of modern cities and living systems according to systems science. We then used a hierarchical cluster analysis method to identify the similarities among the above systems. We analysed the nearest clusters of living systems with city or city components by their major shared traits. Based on our results, we propose a conceptual model that identifies the common features of cities and living systems and that encourages interdisciplinary studies for ecology, biology, planning, and sustainability. Finally, we propose a bionic approach to improve human well-being, reduce environmental pollution, and release more lands for natural recovery.

## RESULTS AND DISCUSSION

**Enclosed industrial systems as components are the symbol of modern cities**

The urban area of a modern city is coupled with peripheral rural areas to form an interdependent self-organizing system (Figure 1A, Figure S1). Modern cities emerged with the rise of enclosed industrial systems, such as factories and chemical plants that produce high fluxes of goods and emit highly concentrated pollutants; newly emerged enclosed ecosystems (such as greenhouses and dairy feedlots) provide food with high efficiency. Of course, modern cities also have many enclosed non-ecosystems provide regulating services and cultural services, such as wastewater treatment plants that purify sewage water, banks that regulate capital, transportation components that regulate the distribution of human goods, and universities that provide education and scientific research.

Both enclosed ecosystems and open artificial ecosystems (such as farmlands) rely on biological processes. For example, greenhouses enhance plant production and livestock feedlots improve animal production, while wastewater treatment plants mainly strengthen the activities of microorganisms (Table S1). The difference between an enclosed ecosystem and an enclosed non-ecosystem is whether or not they have and rely on biological components and processes. For example, a factory or a bank has no biological component except people. It is worth emphasizing that enclosed ecosystems provide food and services which substitutes for natural ecosystems suggesting that increasing enclosed ecosystems can reduce the pressure of land use on nature.

On average globally, enclosed ecosystems now provide 2-5 orders of magnitude greater goods and services per unit area of land than open farmlands and natural ecosystems (Figure 1B and C, Figure S2). This indicates that enclosed ecosystems have become the hotspots of biological goods (food, fibre and water) and ecosystem services (air, microclimate). A wide diversity of enclosed ecosystems has emerged worldwide over the past half century so that now these systems perform basic functions including production (e.g., vegetables, meat, milk and eggs, drinking water), decomposition (e.g., wastewater treatment, waste disposal) and other services (Figure S3, Table S1). In addition, in some regions they provide some goods and services that cannot be provided by open ecosystems. For example, greenhouses can produce fresh vegetables in very cold areas (e.g., high altitude areas in Tibet of China) where open farmlands cannot. Although modern cities grew out of industrialization about 300 years ago, in the early and middle stages of development there were only enclosed non-ecosystems which provide non-ecosystem goods and services[1]. The earliest enclosed ecosystems appeared less than 100 years ago. Having a diversity of enclosed ecosystems is a 'mutation' that helped modern cities step toward maturity because they can produce almost all types of products. The products of enclosed ecosystems can replace foods and other goods and services that are produced by natural ecosystems. Maturing modern cities have characteristics similar to living systems but they are not as similar to organisms as was thought, begging the question, which system type in the hierarchical living system model are modern cities they similar to?

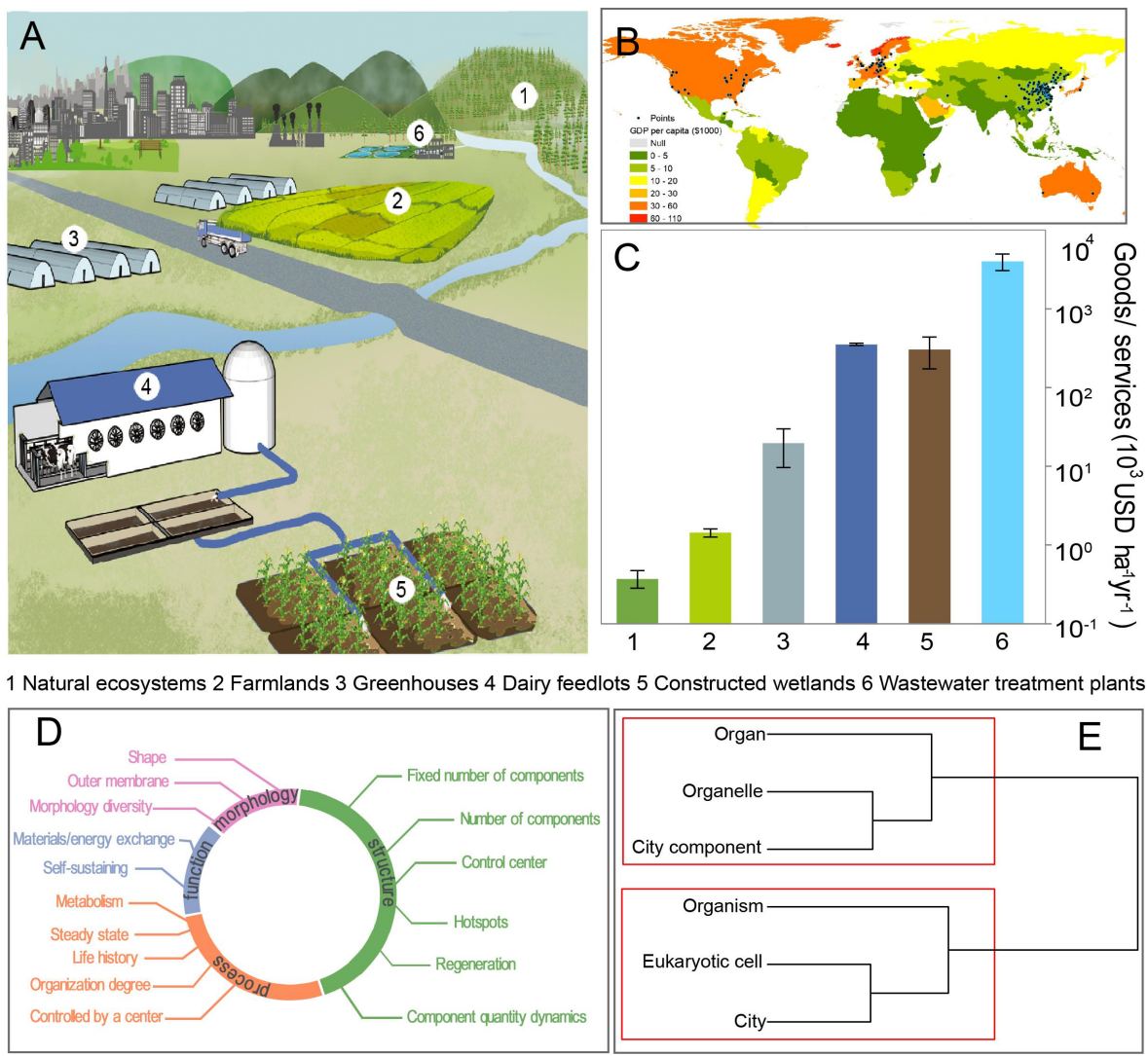

1 Natural ecosystems 2 Farmlands 3 Greenhouses 4 Dairy feedlots 5 Constructed wetlands 6 Wastewater treatment plants

**Figure 1. Some systems and their relationships in hierarchical living systems. (A)** Illustration of some enclosed ecosystems, natural ecosystems and open artificial

ecosystems across an urban-rural landscape; urban green spaces are open ecosystems simulating natural ecosystems. **(B)** Study sites of goods and services of the natural, open and enclosed artificial ecosystems worldwide. **(C)** Goods or services (average of the sample sites), the numbers of the bars correspond to the ecosystem types in Figure 1A. **(D)** The 15 traits of the living systems (data see Table S2). **(E)** Clustering tree of 6 levels of systems (organ, organelle, city functional components, organism, eukaryotic cell and city) based on 15 traits (data see Table S2).

**Similarities between modern cities and eukaryotic cells**

We analysed 15 traits of living systems and cities (Figure 1D, values see Table S2). The clustering results show that modern cities are more similar to the eukaryotic cells than to organisms (Figure 1E). This refutes the super-organism hypothesis and suggests that a new hypothesis should link eukaryotic cells and modern cities (Table 1). In terms of morphology, most cells are circular but a small number are shaped like stars, bars and other irregular shapes. Similar to cells, most cities are a circular or block-like in shape[12] (Figure S1). For the internal structure of a eukaryotic cell, there are thousands of organelles around the cell nucleus (Figure 2A) and each type of organelle is distributed along a spatially distributed pattern on the cell centre-edge. The mitochondria in eukaryotic cells are concentrated near the nucleus while lysosomes are far from the nucleus (Figure 2B), and organelles move in eukaryotic cells (Figure 2C). Similar to eukaryotic cells, there are a great number of enclosed systems around the urban areas in modern cities (Figure 2D). Some enclosed systems, such as banks, restaurants and hotels are concentrated in urban areas; others, such as greenhouses, factories and wastewater treatment plants are located near the urban fringe; while a few, such as dairy farms, are located in exurban areas (Figure 2E). Enclosed systems are also frequently relocated within a city (Figure 2F) either directly, such as railway greenhouses[13], or indirectly, by being dismantled in one place and reconstructed in another. The reason for relocation is the change of relative values for net goods and services of the enclosed systems and because of the changing costs of land leasing due to the development of cities. For example, since the 1980s dairy farms in the Greater Shanghai Area were pushed from the urban fringe to exurban areas many times because they have high negative ecosystem services. The movements of enclosed systems in cities reflect the transformation of the urban development stage, including urbanization (many types of enclosed systems concentrated around the city centre), reverse urbanization (movement outwards from the city centre) and re-urbanization (re-concentration to centre).

**Table 1. Analogies between a eukaryotic cell and a eukarcity**

| Self-organization system | Eukaryotic cell | Eukarcity |
|---|---|---|
| **Functional component** | **Organelle** | **Organara** |
| With outer membrane | Mitochondrion | Greenhouse |
|  | Chloroplast | Dairy feedlot |
| Without outer membrane | Ribosome |  |
|  | Centrosome |  |
| With information storage | Chloroplast | Dairy feedlot |
|  | Mitochondrion | Wastewater treatment plant |
| Without information storage | Ribosome | Greenhouse |
|  | Golgi apparatus |  |
| **Nucleus** | **Cell nucleus** | **Citynucleus** |
| Information storage | DNA | Stored written information |
| Regulator | RNA | Active written/oral information |
| Nuclear matrix |  | Green space |
| **Matrix** | **Cytoplasm** | **Cityplasm** |
|  | Soluble enzymes | Forests |
|  | Sugars | Grasslands |
|  | Inorganic salts | Aquatic ecosystems |

From the perspective of physics, the number of chloroplast and mitochondria, in response to cell size (volume) is sublinear, $β = 0.51$ and $0.53$, respectively (Figure 2G). Similar allometric scale effects are also found in modern cities: the number of gasoline stations in response to city size (population) is $β = 0.84$ in China (Figure 2H), 0.77 in United State of America[6]; and for wastewater treatment plants with an average $β = 0.77$ among China, United State of America, France and Germany (Figure 2I). In contrast, the number of organs in an organism is often fixed, so there are no such scale effects in response to the increase of body size. The allometric scale effect of components in response to system size is a general principle for both cities and eukaryotic cells.

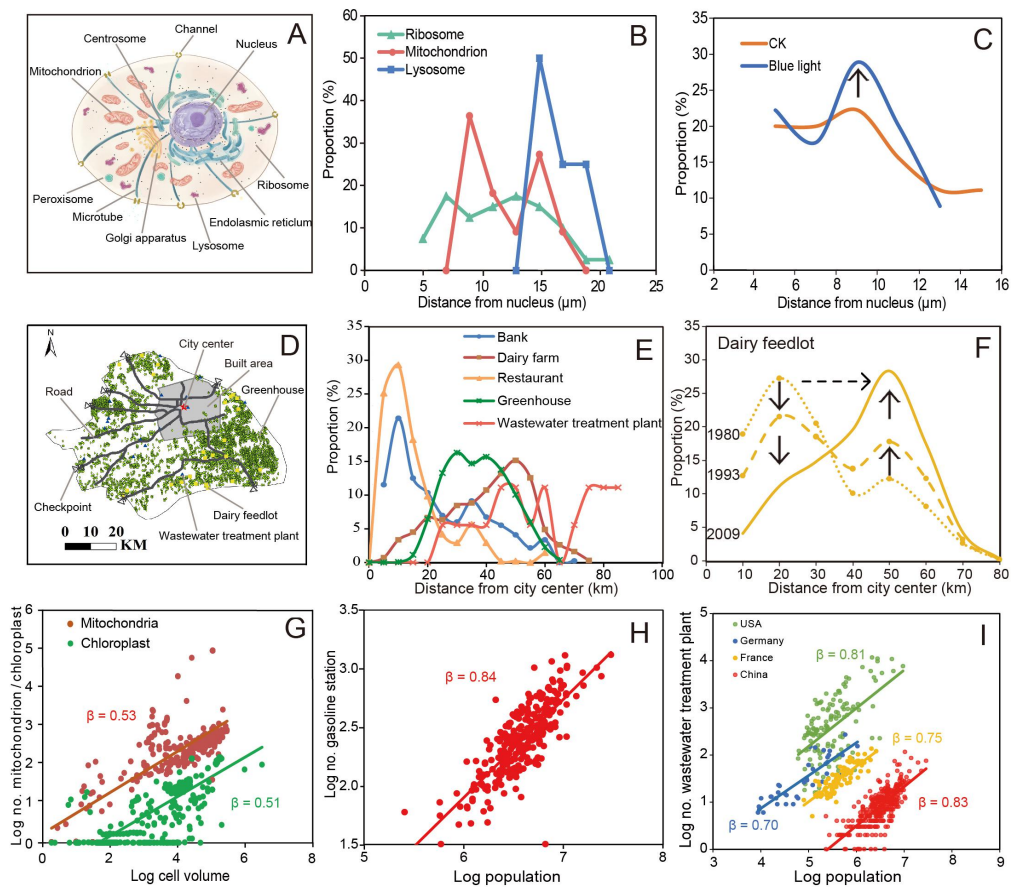

**Figure 2. Morphology and substructure of modern cities and eukaryotic cells. (A)** An animal cell with nucleus and microtubules meeting, organelles and cytoplasm as matrix; **(B)** distribution of some organelles along the nucleus-edge gradients in a plant cell; **(C)** movement of chloroplasts under blue light (CK is the normal state); **(D)** Greater Shanghai Area, with a city core and several roads and enclosed ecosystems as city-components and other ecosystems as matrix of landscape; **(E)** density of enclosed ecosystems along the urban-rural gradient in Shanghai in 2010; **(F)** distribution patterns of dairy feedlots in Shanghai from 1980 to 2009; **(G)** the quantity of organelles (chloroplasts and mitochondria) in response to the cell volumes (scale effect, $Y = aX^\beta$); **(H)** the quantity of city greenhouses in response to the population size in cities of China; **(I)** wastewater treatment plants in response to the population size in cities. Data set see Supplementary Materials.

The metabolic flow and network characteristics of the metabolisms of modern cities are also similar to those of eukaryotic cells. In eukaryotic cells, organelles are the hotspots of metabolism[14]; whereas in cities, enclosed systems are the hotspots of biogeochemical metabolism. For example, nitrogen fluxes passing through enclosed systems are up to three orders of magnitude higher on average than those in open systems (Figure S4, A to C). The nitrogen flux of the nodes in natural ecosystem food web is low, and the attenuation rates (power) of nitrogen fluxes ($|\beta| < 2$) are much lower than the power in the nitrogen network of cities ($|\beta| > 3$). This reveals that nitrogen fluxes in cities are centralized in some hotspots, all of which are enclosed systems, while in natural ecosystems the fluxes are less centralized (Figure S4D). Such a high centralization of metabolic flows in a few hotspots components illuminates the similarities in biogeochemical cycle patterns between cities and eukaryotic cells. It suggests that metabolism patterns in modern cities are more like those in eukaryotic cells rather than organisms.

The nonlinear scale effect in metabolisms is due to the self-organization in cells and cities. The internal integration of the components within the system is reflected in the nonlinear response to the external changes[6]. The outer membranes of cells moderately isolate the interior and the exterior and they selectively exchange materials[15], energy, and information with the external environment to ensure the self-organization. Cities do not have outer membranes like eukaryotic cells but do have administrative boundaries on the fringe of the urban-rural systems (Figure 3A, Figure S1). In order to perform selective permeability, there are control devices (such as roads and harbour checkpoints) on the boundaries of cities to allow the entry of goods and positive services, but to prevent harmful goods (e.g., garbage and wastewater) from outside[16]. In sum, because of the analogies that can be drawn between modern cities and eukaryotic cells in structure, metabolism and self-organization, we henceforth refer to a modern city as a 'eukaryotic city' or a 'eukarcity'. The new term 'eukarcity' corrects the type relationship of modern cities in the hierarchical living system and lays a solid foundation for scaling up and down with eukaryotic cells.

**Similarities between enclosed industrial systems and organelles**

The cluster analysis also demonstrates another closely related pairing (Figure 1E): enclosed systems are similar to organelles. Many organelles in eukaryotic cells (such as chloroplasts and mitochondria) have outer membranes to maintain their physical and chemical homeostasis; similarly, enclosed industrial systems have outer membranes to ensure the stability of indoor physical and chemical conditions (Figure 3A, Table 1). For example, air temperature variations within greenhouses, vertical farms, and dairy feedlots are much smaller than ambient environmental variations (Figure 3B to E). That is why greenhouses can be used in extreme climate zones like Tibet to produce food under conditions that were previously impossible. In addition to maintaining homeostasis that strengthens resistance to climate variations, membranes also play an important role in reducing the release of 'intermediates' (such as ketones, metal ions) and polluting the cytoplasm. Similar to this, the outer membranes of enclosed industrial systems are gradually improved to mitigate pollution. More importantly, organelle membranes have a great number of small 'facilities'[15], such as channels, ion pumps, glycoproteins, ATPase, and receptors (Figure 3A top-left insert). Similarly, the outer membranes of enclosed industrial systems have been gradually outfitted with small facilities such as sensors, monitors, air-conditioners, solar batteries, and so on (Figure 3A top-left insert), and are rapidly increasing in quantity. It can be expected that the development of these fine structures will greatly improve the functions of the enclosed system.

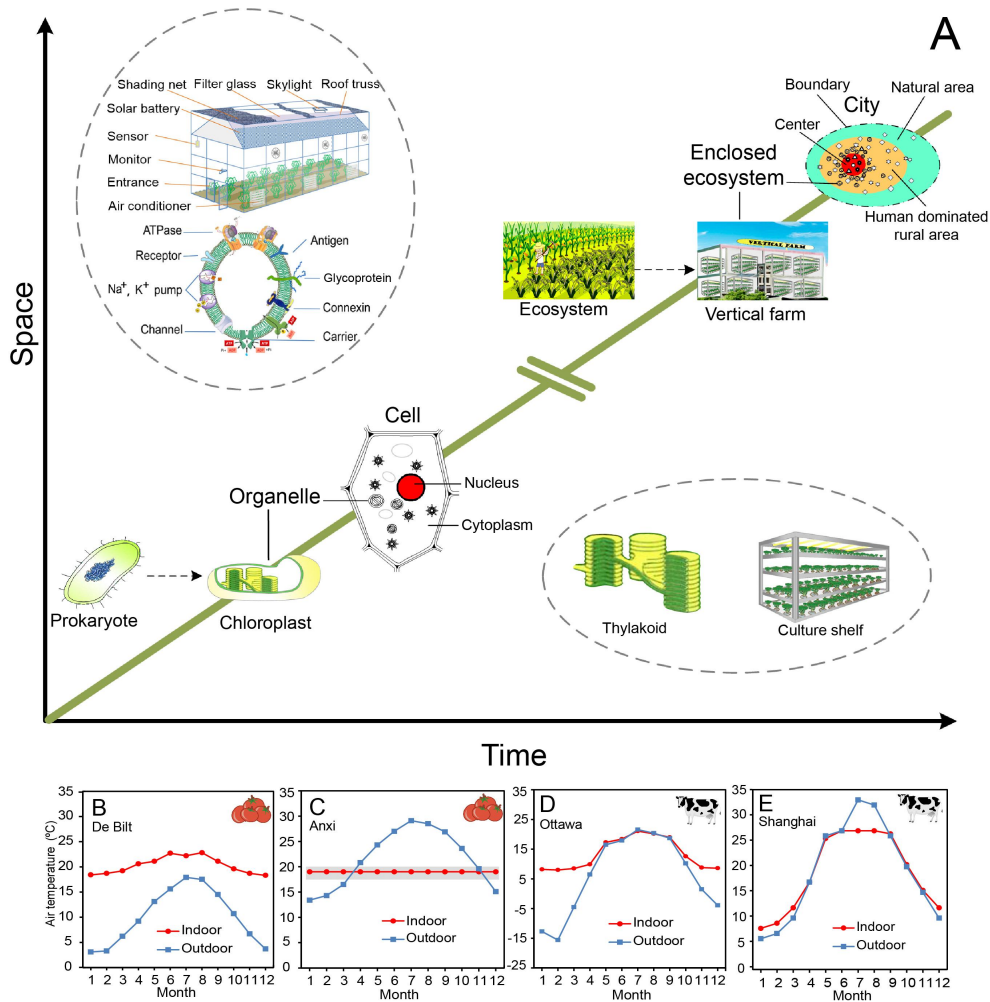

**Figure 3. Similarities of structure and homeostasis between a eukaryotic cell and a modern city.** **(A)** Hierarchical systems in time and space. The discontinuous labeling on the bold green lines denotes that enclosed ecosystems and cities are much greater in size than eukaryotic cells, and that a city is consist of urban area (red) and human-dominated rural areas (orange), dash-dotted line denotes the administrative boundary; the top-left inset indicates the structural similarities between the outer membrane of an enclosed ecosystem (e.g., glass greenhouse) and an organelle (e.g., chloroplast); the bottom-right inset indicates the relationship between the internal structure of the thylakoid and multi-layer planting system; the dashed circles do not relate to the X-axis or Y-axis. (**B to E**) Variations of internal and external air temperature of enclosed ecosystems, including greenhouses in De Bilt, the Netherlands (**B**), vertical farm (crop factories) in Anxi, South China (**C**), dairy feedlots in Ottawa, Canada (**D**), and dairy feedlots in Shanghai, Southeast China (**E**).

The internal structures of enclosed systems have also developed some fine structures similar to organelles (Figure 3A bottom-right insert). For example, a chloroplast has a multi-layer structure to improve photosynthetic efficiency[17]; similarly, a vertical farm uses multi-layer planting to centralize cultivation with 20 to 100 tiers, which involves much technology and automation to achieve high yields throughout the year[18]. Similar to the fine structure formed by the inner membrane of mitochondrion[19], a cowshed has many stalls which efficiently uses space and avoids crowding with a trough configuration that gives each cow an equal chance of getting feed. These fine structures can increase the efficiency of feed utilization and dairy productivity (Figure S3). Yet, the fine structures are barely developed compared with those in organelles suggesting possibilities for further evolution.

Another similarity between enclosed systems and organelles lies in their information systems. Enclosed non-ecosystems (factories) have only human information systems, such as technology, management, and standards, while enclosed ecosystems have dual-information systems, i.e., human information and biological information. For example, industrial dairy feedlots have biological information systems including the age, gender, and genetic structures of cattle populations. The information collected in enclosed industrial systems ensures their self-organization and is similar to some organelles' information systems[20]. Within a eukaryotic cell, the organelles and nucleus interact frequently as enclosed systems interact with city core through human information exchange.

The analogies between enclosed systems and organelles discussed above suggest that we can define an enclosed system as an 'organara'. The term organara follows organelles in terminology for the suffix '-*elle*' in Greek means small, while '-*ara*' means big. The emergence of organaras has led to the transformation of traditional cities to eukarcities, just as organelles significantly transformed prokaryotic cells to eukaryotic cells. The new concept of organaras lays the foundation for the construction of a new city model which is described below.

**A conceptual super-cell city model**

The results of our analysis encouraged us to develop a new hypothesis — a modern city is analogous to a eukaryotic cell but not to an organism (Figure 3A). Modern cities mimic eukaryotic cells in their structure, organization, and functions; organaras serve the function of organelles in cities as mentioned above. We therefore propose a 'super-cell city' model (Figure 4). The model is based on the super-cell city hypothesis, coupling the flow of artificial goods, services and information based on organaras as nodes and the flow of energy, biomass and information based on ecosystems. We define organaras as a new type of quasi-living system (Figure 3A; Figure S5), which correspond to organelles but at a higher level in hierarchical living system. We suggest that eukarcities and organaras are the latest milestones of evolution on Earth and we provide further evidence below.

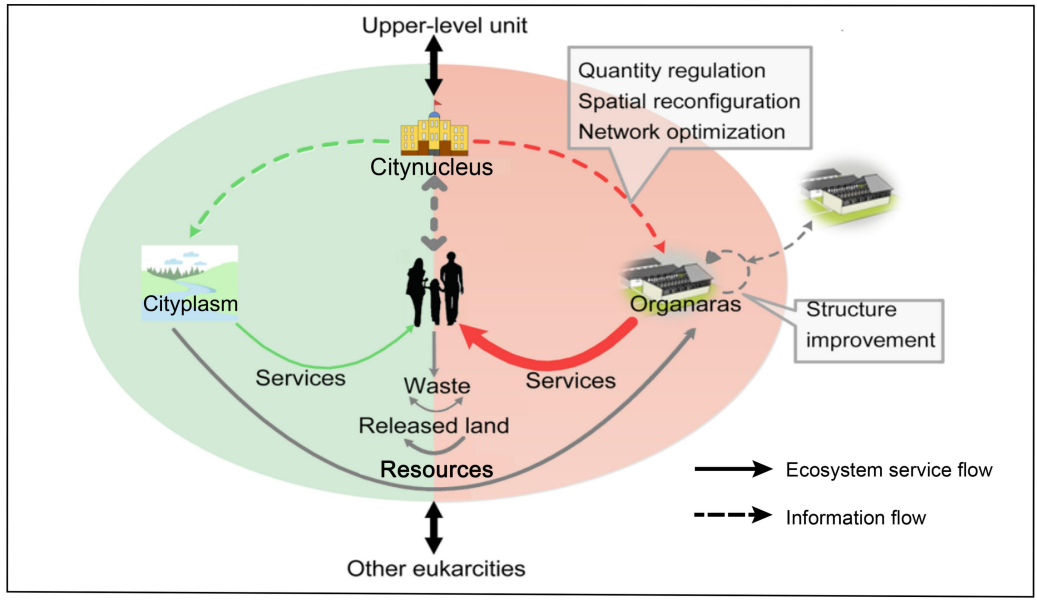

**Figure 4. Illustration of the super-cell city model.** A eukaryotic city composed of the processes dominated by cityplasm (left) and the processes dominate by organaras (right). Thickness of lines indicates the intensity of ecosystem services; interactions between organaras and cityplasm are mainly by means of land use change, natural capital, waste production and treatment. Organaras always provide more ecosystem services than nature, so the right semicircle plays an increasing role in eukarcities. A eukarcity interacts with other eukarcities though flows of goods and services, as well as information. Organaras interact with other organaras (often the same type) within the same eukarcities or in other eukarcities though information.

The 'citynucleus' of a eukarcity is located in the urban area. The citynucleus regulates the whole city and integrates urban and rural areas through policies, science, technology, culture, market and finance, including the number of organaras and their spatial distribution and networks (Figure 2D). Organaras provide major goods, such as food production (corresponding to 'synthesis' in organelles), and services, such as garbage disposal (corresponding to 'degradation' in organelles), to residents. The 'cityplasm' is made up of the natural ecosystems and open artificial ecosystems (Figure 3A). In addition to the goods provided by farmlands, the cityplasm provides regulating and supporting services such as maintaining air quality, water cycling, soil and biodiversity. Compared to organaras, the cityplasm is characterized by low fluxes of energy, material and productivity. The quality of the cityplasm, such as quality of air, water bodies, soil, and ecosystem health is mainly affected by organaras. For example, restaurants and dairy feedlots discharge wastewater and cause water pollution[21]. In contrast, there was almost no water or air pollution in pre-industrial traditional cities. Now the cityplasm is frequently monitored by people and the data collected are fed back to the citynucleus, which then modifies the institutions and policies for optimizing the city through regulating organaras as units. Eukarcities and organaras also have many 'grey' infrastructures (such as cement, metal, and glass) (Table S3) that are non-living materials just as eukaryotic cells have non-living materials including cytoskeletons, cell walls, and vacuoles suggesting that non-living components are a common feature of living systems.

There is much evidence that supports the utility of the super-cell city model in urban design and management. One successful case occurred in Lake Taihu in southeast China where over the past 30 years water quality deteriorated and then recovered due to increased organaras (Figure 5, A to C). The lake relied on its abundant wetlands for wastewater purification for over past 3,000 years. After the 1980s, however, industrial waste and farmland fertilization exceeded the purification capacities of the natural ecosystems and water quality rapidly degraded. In Wuxi city, where the drinking water comes from Taihu, a famous drinking water interruption event occurred in 2007 when the water became non-potable. Since the economic upturn (Figure 5D), the number of wastewater treatment plants in all cities (Figure S1A) around Lake Taihu increased from 5 in 1985 to 331 in 2014 (Figure 5E), and water quality improved even though the population is increasing continuously (Figure 5F). This case clearly illustrates that decomposition-type organaras have been crucial effective functional components for mitigating environmental pollutions in modern cities.

Some evidence supports the view that industrialization of ecosystems increases the supply of goods and services and reduces the ecological footprint per unit product[22]. In China, decreasing farmland area has corresponded with sharp increases in the number of greenhouses (Figure 5G). The area of sustainable pasture, which maintains nearly natural community structure, is decreasing in productivity while the number of industrial dairy feedlots which provide the major of dairy products in China is increasing quickly (Figure 5H), relieving the stress on pastures. In contrast to decreasing farmlands and grasslands, forests and wetlands in China are continually being restored as economic development progresses (Figure 5I). These examples confirm that the development of organaras has decreased the human footprint, at least in China. It also suggests that the super-cell city model can guide future cities to reduce pollution, optimize spatial structure and release land for natural recovery by consciously regulating organaras.

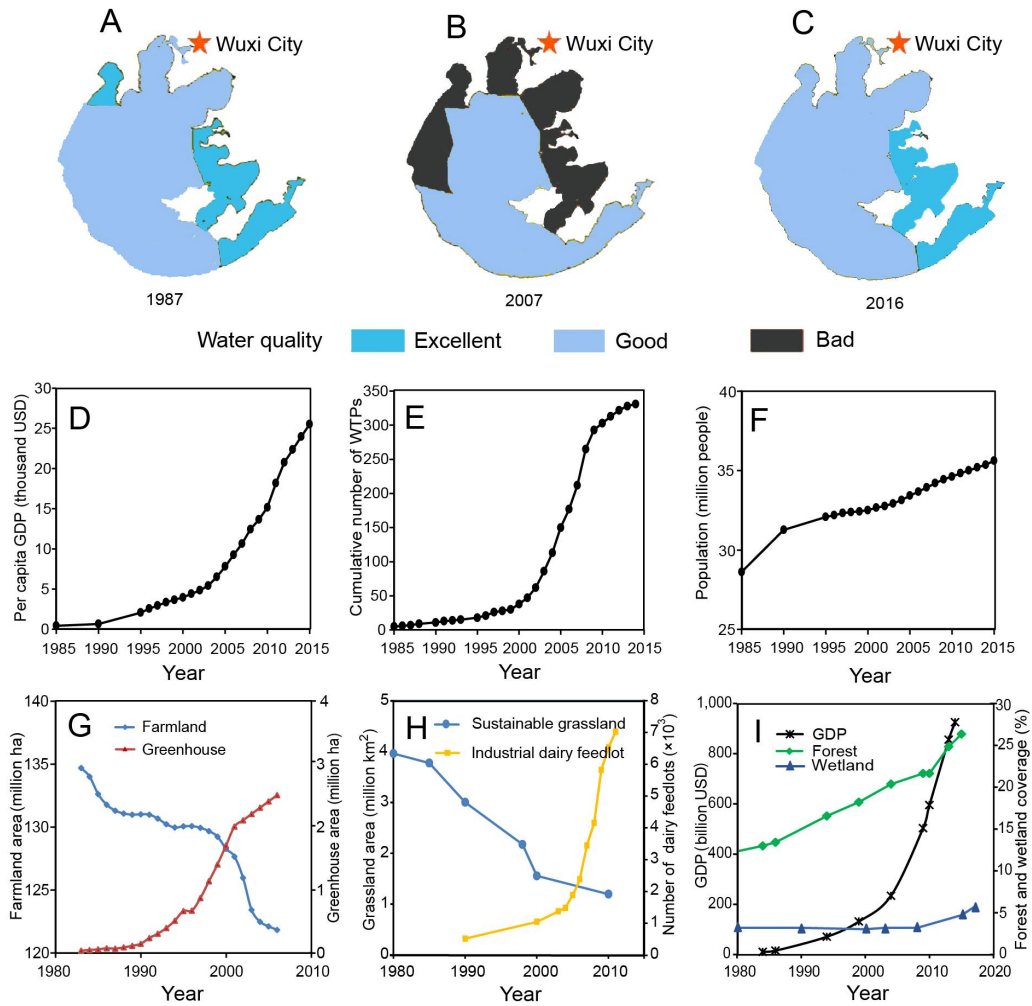

**Figure 5. Historical trends in the amounts of organaras and cityplasm in China from the 1980s to 2010s.** (**A to C**) Water quality of Lake Taihu, southeast China in 1987, 2007 and 2016, respectively; (**D to F**) historical change of the social-ecological state in Lake Taihu area, including per capita GDP (**D**), annual increment of wastewater treatment plants (WTPs) (**E**) and population (**F**); (**G**) decrease of open farmland area and increase of greenhouses; (**H**) sharp decrease of sustainable grassland area and increase of industrial dairy feedlots; (**I**) synergy between economic development and forest restoration.

**Super-cell city model makes hierarchical living system seamless in scale and principle**

A hierarchical living system model has been built from the bio-macromolecular to the entire globe, yet the model left a sizable gap of several orders of magnitude between ecosystems (~$10^3$ m in size) and the biosphere (~$10^8$ m)[23]. The super-organism hypothesis [6,24] put 'city' into the hierarchical model which is helpful in terms of filling the gap because cities (urban-rural system, average ~$10^5$ m in size) fall just within the gap of the original hierarchical living system model (from $10^3$ m to $10^8$ m). However, based on the evidence that we have presented here, we suggest that the super-cell city model should replace the super-organism model. Furthermore, the city as the counterpart of a eukaryotic cell should not be limited to an urban area, but is the urban area also coupled with human-dominated rural area. The emergence of organaras forms the basis for eukarcities just as organelles are the basis of eukaryotic cells revealing that the shared features between cities and cells can be scaling up and down on the hierarchical living system. We further suggest that, following the origin of life, the emergence of eukaryotic life[25,26] and the appearance of human beings, the emergence of the eukarcity is the fourth evolution event.

**Super-cell city model extending endosymbiosis theory**

The super-cell city model extends endosymbiosis theory from eukaryotic cells to the city level. According to endosymbiosis theory, eukaryotic cells were a 'big vacant cells' that swallowed prokaryotic cells, such as cyanobacteria and spirochetes, to form a symbiotic fusion[26]. There is evidence that some organelles retained their own genes which are able to duplicate themselves and interact with the genes in the nucleus, to achieve semi-autonomous regulation[20]. For example, chloroplasts and mitochondria — besides being controlled by nuclear genes — also have their own DNA[26]. Inspired by endosymbiosis theory, we hypothesize that the origin of the eukarcity is a symbiotic fusion of organaras (enclosed ecosystems and enclosed non-ecosystems) with pre-industrial traditional cities. Like organelles, some organaras are also semi-autonomous in that in addition to being controlled by the citynucleus, they also have their own "genetic information" for duplication and operation[16]. For example, the decomposition of pollutants in wastewater treatment plants is carried out by microorganisms which have gene pools; besides, there are human information including libraries, laboratories and offices that containing the technology, records, and management files (Figure S3). This 'cultural gene' can be interpreted, copied, transmitted and modified, and constantly innovated. The super-cell city model adopts endosymbiosis theory to explain the evolution of the modern city and also provides a theoretical basis for city research, design and management.

**Super-cell city model helps in developing new management paradigm**

The new principles revealed by the super-cell city model also require us to find new rules to adapt to the new systems, from the natural ecosystems management paradigm to cities management paradigm. In other words, modern cities should learn from eukaryotic cells and organelles by taking organaras as their operating units in order to optimize urban structures, processes and functions. This approach mainly includes: (1) Following the optimized principles of organelles in terms of structure and functioning[19], improving organaras by applying advanced technologies (including artificial intelligence). (2) Identifying the types, quantities and spatial distribution patterns of organaras in cities, diagnosing existing problems and improving the city functional and spatial structure by learning the optimized principles of organelles in eukaryotic cells. (3) Improving supply efficiency and robustness by regulating organaras according to the high metabolic efficiency and robustness of eukaryotic cells[27]. The principles described above belong to the bionic approach which is generally superior to 'trial and error' methods[28] because they are guided by natural selection and can reduce trial times while increasing efficiency. In sum, the super-cell city model proposes that cities can learn from eukaryotic cells in traits and principles which have been optimized via evolution over billions of years.

The weakest aspect of organaras currently is structural improvement. At present, the outer membranes and internal structures of organaras are very preliminary relative to the complex structure and important functions of organelle membranes[29], suggesting a great potential for technological improvements in organara membranes structure. For the outer membranes, some case studies have found that the heavy glycosylation of proteins located on the lysosomal membrane protects the membrane from degradation by lysosomal hydrolases[30], and avoids the release of hydrolase from lysosome to cytoplasm causing damage to cells. Cyclochrome P450 proteins are distributed in the membrane of the endoplasmic reticulum, mitochondria, Golgi apparatus and other organelles, and have the function of detoxification[31], and can prevent toxins from leaking into the cytoplasm. Organaras can learn from these mechanisms to improve outer membranes. For example, air filters installed in livestock farms can prevent virus entry[32].

For the internal structure to be advanced, the mitochondrial inner membrane is divided into distinct regions which are not only morphologically distinct, but also appear to constitute separate functional domains[19]. Likewise, each thylakoid stacked inside a chloroplast is an enclosed unit that improves the efficiency of photosynthesis[17]. Inspired by these traits, organaras can also be compartmentalized by internal membranes into specialized subunits to increase efficiency. Thus, it can be expected that increasing the structural similarities between organaras and organelles will greatly improve organara supply capacity of goods and services. Furthermore, making organaras even more enclosed would make them even less influenced by climate; thus, the eukarcities in different regions can adopt the appropriate strategies for efficiency and productivity.

**Implications of the model**

It is ironic that human consumption of goods and services is dramatically increasing while many natural ecosystems are already degraded and severely reduced in size and thus are providing fewer and fewer goods and services[5]. According to the super-cell city model, the structural improvements of organaras and eukarcities may address this conflict. Our scenario analysis shows that in Shanghai when the ecosystem services provided by 1 hectare of greenhouse are increased by 10% through structural improvement, the added services will be equivalent to those provided by 5.5 hectares of open farmland; if a 1 hectare of wastewater treatment plant increases its services by 10%, it is equivalent to an increase of 5.3 hectares of constructed wetlands or 371,350 hectares of natural wetlands. This means that much land would be released if the open artificial ecosystems can be replaced by organaras. The marginal benefit of technology is that these new goods and services occupy less land thus releasing the land occupied by open artificial ecosystems to achieve a win-win of economic development and ecosystem recovery.

Of course, organaras increase environmental pollution intensity. For example, the intense industrialization of dairy feedlots increases the pollution intensity per unit area of land and increases ecological risk[33]. Therefore, high intensity pollution reduction approaches need to be explored. Our previous study found that the coupling of decomposition organara and production organara is a potential way to solve this kind of problem. For example, coupling dairy feedlots with constructed wetlands can reuse waste nutrients and greatly mitigate pollution[21]. The nutrient recycling mode between the organaras can improve yield and reduce the pollution, which consistent with the optimization mode of eukaryotic cells.

Furthermore, organaras centralize metabolic fluxes and can increase efficiency, but also decrease metabolic robustness due to the lack of alternative pathways. Theoretically, increasing parallel pathways in a network can make it more robust[28]. Hence, it has to improve capacity and strength of eukarcities by increasing parallel pathways of processes by increasing the number of organaras and optimizing the networks after eukarcity goods and services production efficiency has been improved. Of course, these inputs increase the cost of cities, so this approach requires policy makers to identify the trade-offs in decision-making with respect to the best solutions under a specific set of conditions for social-ecological systems. At present, there have been attempts to use artificial intelligence (AI) to organaras (such as AI-livestock farm, AI-vertical farm and AI-restaurant) in cities with high economic level[34]. The innovations with intelligent organaras as nodes of metabolic network will promote city's physical structure and improve city's vitality. In sum, the super-cell city model is helpful for cities to adapt the most advanced technology to complement organaras, as well as the network optimization of entire cities.

Although organaras are increasing in number and expanding in area, most open farmlands and pastures will not be converted into organaras, just as most of the space in eukaryotic cells is still cytoplasm[35]. The cityplasm will continue to provide some key ecosystem services and support organaras in future eukarcities. Certainly, some goods and services, such as fresh water, wildlife habitats, biodiversity and some cultural services can only be provided by natural ecosystems[16,36]. Likewise, the natural capital from the cityplasm will continue to support the production of the goods and services of organaras. For example, open farmlands, pastures and forests will continue to provide grain, fibre and raw materials. Therefore, the super-cell city model requires serious collaborative management between natural ecosystems and organaras in order to realize city sustainability.

**CONCLUSION**

The super-cell city model provides a solid basis in principle and in methodology for finding the shared features among living systems and two levels of quasi-living systems (eukarcities and organaras). The model is a major theoretical step linking cities and cells and promotes new knowledge. In-depth research of this concept model, such as additional theoretical and mathematical models, will be the intriguing challenges. The super-cell city model encourages transdisciplinary studies and education on cities and cells, and connects multiple disciplines, including engineering, science, policy, and culture in a coherent manner to conduct this in-depth research and practice. The rapidly urbanizing world can provide much empirical evidence for such studies. The analysis in this article is just the tip of the iceberg and our purpose here is not to detail every aspect but instead to introduce a methodology for analysis the new feature of the changing world. In the future, every nation's governance will likely be based on cities. We suggest that the super-cell city model not only contributes to the development of cities, but also contributes to global development by guiding all cities towards supporting a vibrant and sustainable world.

**EXPERIMENTAL PROCEDURES**

**Trait selection**

*The quantitative and qualitative traits.* We chose the shared features of 6 types of living systems, including modern cities, enclosed industrial systems, organisms, organs, eukaryotic cells and organelles. According to systems science, we identified the traits related to morphology, structure, process and function of the systems (Table S2). The 15 traits include 7 quantitative traits and 8 qualitative traits. For the quantitative traits, some traits can be described with an accurate value, and some values are expressed as an order of magnitude; other traits are the power ($β$) of power law to express the quantity of components or organelles, or metabolic rate in response to the system size. For the qualitative traits, we assigned the attribute values (detail see Table S2).

*Definition of some ecosystems not included as the enclosed ecosystem.* Besides the enclosed ecosystems (Table S1), farmlands and rangelands are open artificial ecosystems. In particular, even farmland areas using plastic film mulch are not considered to be organaras because their internal structure is too simple and the cover only works during the early period of plant growth. Constructed wetlands are not organaras even though they have a waterproof liner on the bottom, and they have no top cover to maintain stable internal conditions.

*Cluster analysis.* We used a hierarchical cluster analysis with the complete linkage method based on Gower distance to identify relatedness among living systems.

**Assessment of ecosystem services**

*Frameworks* Ecosystem services include provisioning services, regulating services, and cultural services. It should be noted that the calculation of the ecosystem services of natural ecosystems and urban green spaces are the sum of all three services (provisioning, regulating and cultural) provided. However, the ecosystem services of artificial ecosystems are divided into target services (provisioning or regulating services) and accompanying services separately. For example, the target service of farmlands, greenhouses, and dairy feedlots is food production, which is equivalent to some of the provisioning services of natural ecosystems; the target service of wastewater treatment plants and constructed wetlands is pollutant removal from wastewater, which is equivalent to the regulating services of natural wetlands. In addition to the target services, the calculation of accompanying services is the same if they are provisioning, regulating or cultural services. Regulating services are further divided into services (positive) and disservices (negative) in this paper, following the guidelines in Liu et al.[37].

*Data collection* Greenhouses were identified visually and marked using Google Earth™. The locations of wastewater treatment plants and dairy feedlots were searched from Baidu Maps™ and then marked using Google Earth™. To ensure the reliability and accuracy of the locations, we sampled some of the sites when conducting the field survey and recorded the locations by handheld GPS. For the number of roads in a city, we visualized the images using Google Earth™ at an appropriate scale that could clearly distinguish the main highways that surrounded the city, and we then counted the number of the main highways that link the city with other cities. Some data are collected from the literature. For details, see the Supplementary Materials.

*Calculations* Ecosystem services for natural ecosystems and urban green spaces were calculated as described in Chang et al.[38]. The provisioning services ($ES_p$) of greenhouses or open farmlands were calculated as described in Chang et al.[39]. The provisioning services ($ES_p$) of dairy feedlots were calculated as described in Fan et al.[21]. The decomposition services ($ES_d$, USD ha$^{-1}$ yr$^{-1}$) of wastewater treatment plants, landfills and waste incineration plants, including those providing solid waste or wastewater treatment were calculated in this study as follow:

$$ES_d = W \times P / A$$

where $W$ (ton ha$^{-1}$ yr$^{-1}$) is the amount of waste treated at wastewater treatment plants, landfills or waste incineration plants, $P$ ($ ton$^{-1}$) is price for per unit of waste treated, and $A$ (ha) is the area covered by wastewater treatment plants, landfills or waste incineration plants in a city.

**Assessment of the scale effect (scaling law)**

We chose greenhouses to represent organaras for food production services and wastewater treatment plants to represent decomposition organaras in cities in order to calculate the scaling effect. The scaling effects are the power law functions ($Y = aX^\beta$) between the number of a type of organara (Y) in response of the population (X) in the city. The method follows Bettencourt[4].

## SUPPLEMENTAL INFORMATION

Supplemental Information can be found online at https://

## ACKNOWLEDGMENTS

This work was financially supported by the National Science Foundation of China (Grant No. 31870307, 31770434, 31670329). We thank XZ Liu, RH. Xu, WJ. Han, LF. Shang, SN. He, B. Luo, GY. Luo, LS. Lin, X. Wang, Q. Wang, YY. Duan, WL. Gu and Y. Chen S. Liu, for the data collection and manuscript checking; Y. Geng, SD. Niu for their comments.

## AUTHOR CONTRIBUTIONS

J.C. and Y.G. conceived the study and wrote the first draft of the paper, L.A.M. and M.P.H. contributed to further formulation of the ideas. Z.W., Y.D., K.P., G.Y., Y.R., F.M., Z.Q., X.F., Y.M., C.P. contributed to data collection, calculation and discussions and revisions.

## DECLARATION OF INTERESTS

The authors declare no competing financial interests.

# Supplemental Information

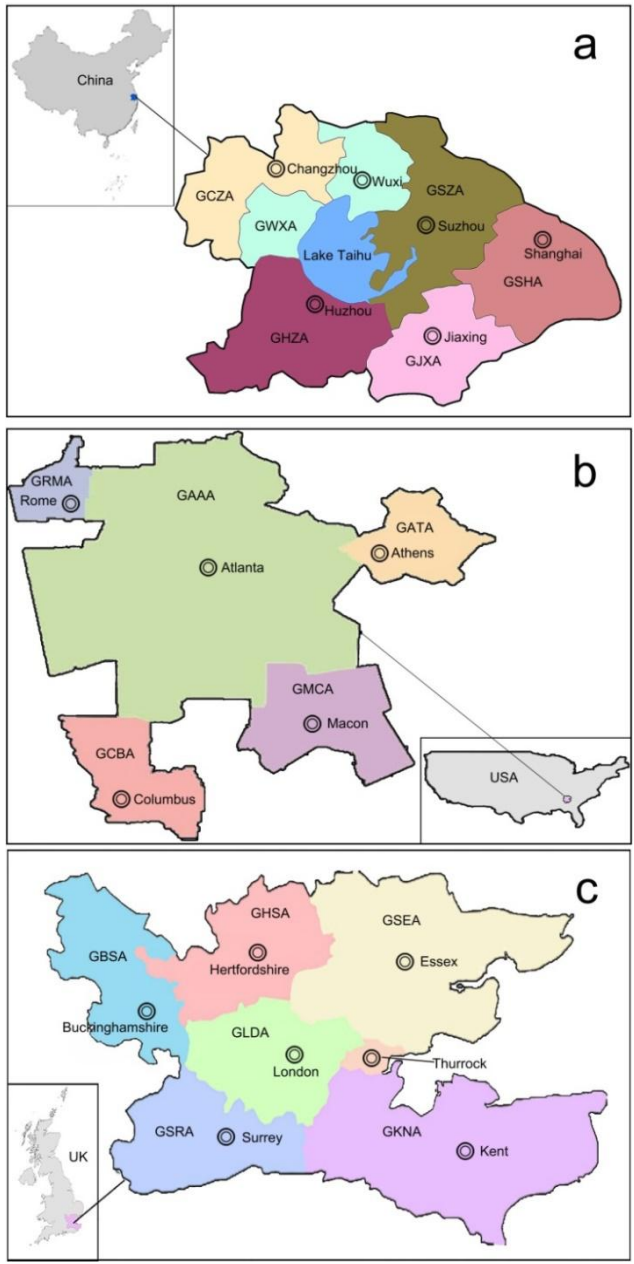

**Figure S1. Urban areas and rural areas, in China, the United States, and United Kingdom. a.** Cities around Lake Taihu, Southeast China: Greater Shanghai area (GSHA), Greater Jiaxing area (GJXA), Greater Huzhou area (GHZA), Greater Suzhou area (GSZA), Greater Wuxi area (GWXA), and Greater Changzhou area (GCZA), and each city has the administrative boundary with others; **b.** Cities in Southeast USA: Greater Atlanta area (GAAA), Greater Athens area (GATA), Greater Macon area (GMCA), Greater Rome area (GRMA), and Greater Columbus area (GCBA); **c.** Cities in southern United Kingdom: Greater Buckinghamshire area (GBSA), Greater Hertfordshire area (GHSA), Greater Essex area (GEA), Greater London area (GLDA), Greater Thurrock area (GTRA), Greater Surrey area (GSRA), and Greater Kent area (GKNA).

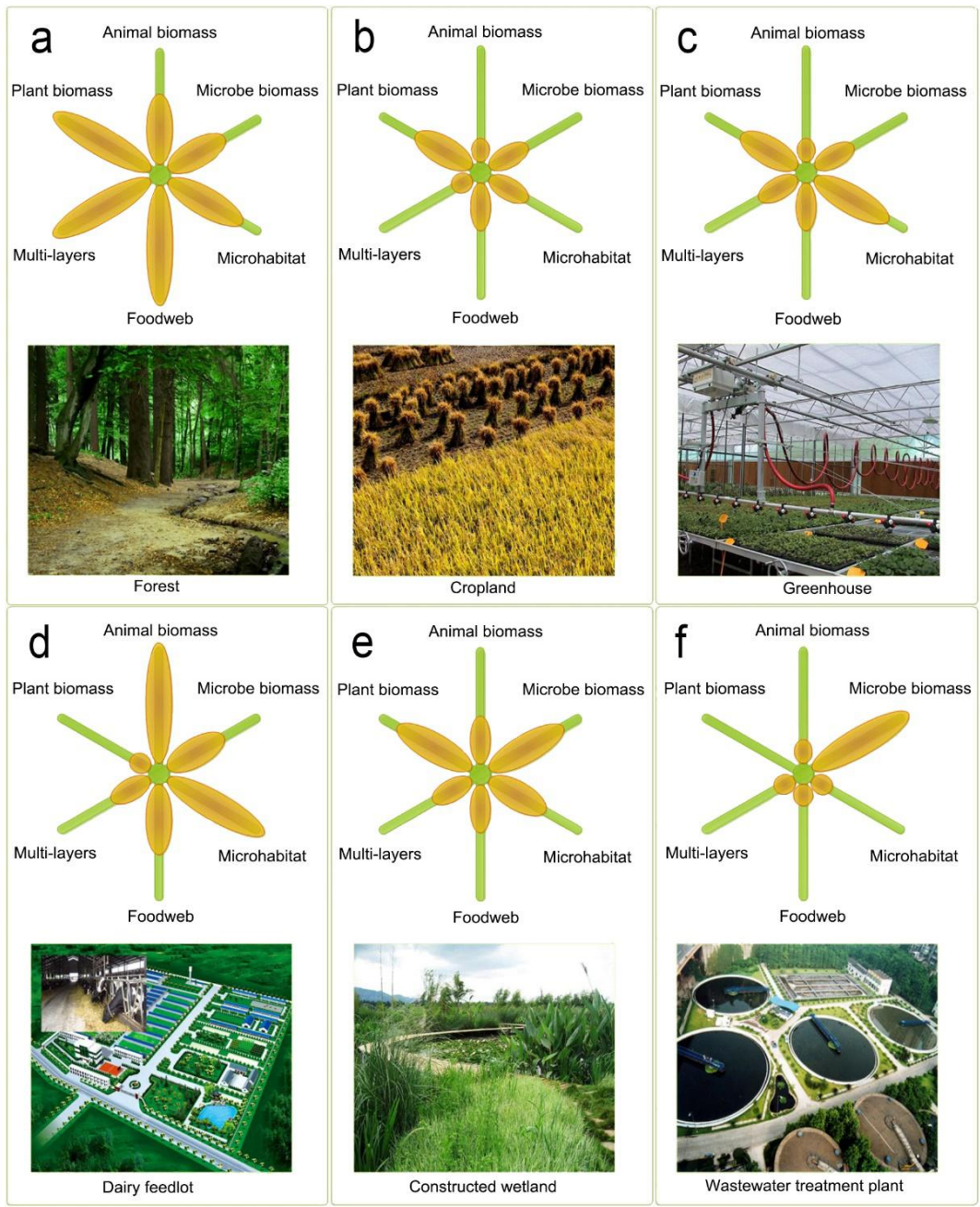

**Figure S2. Illustrations of natural ecosystems, open artificial ecosystems and those have transformed to enclosed ecosystems.** The original natural ecosystem structure was simplified and reconstructed by human beings. Compared with forests (representing a complex natural ecosystem), open farmlands, greenhouses, dairy feedlots, constructed wetlands and wastewater treatment plants are weakened or have lost many of their major natural features.

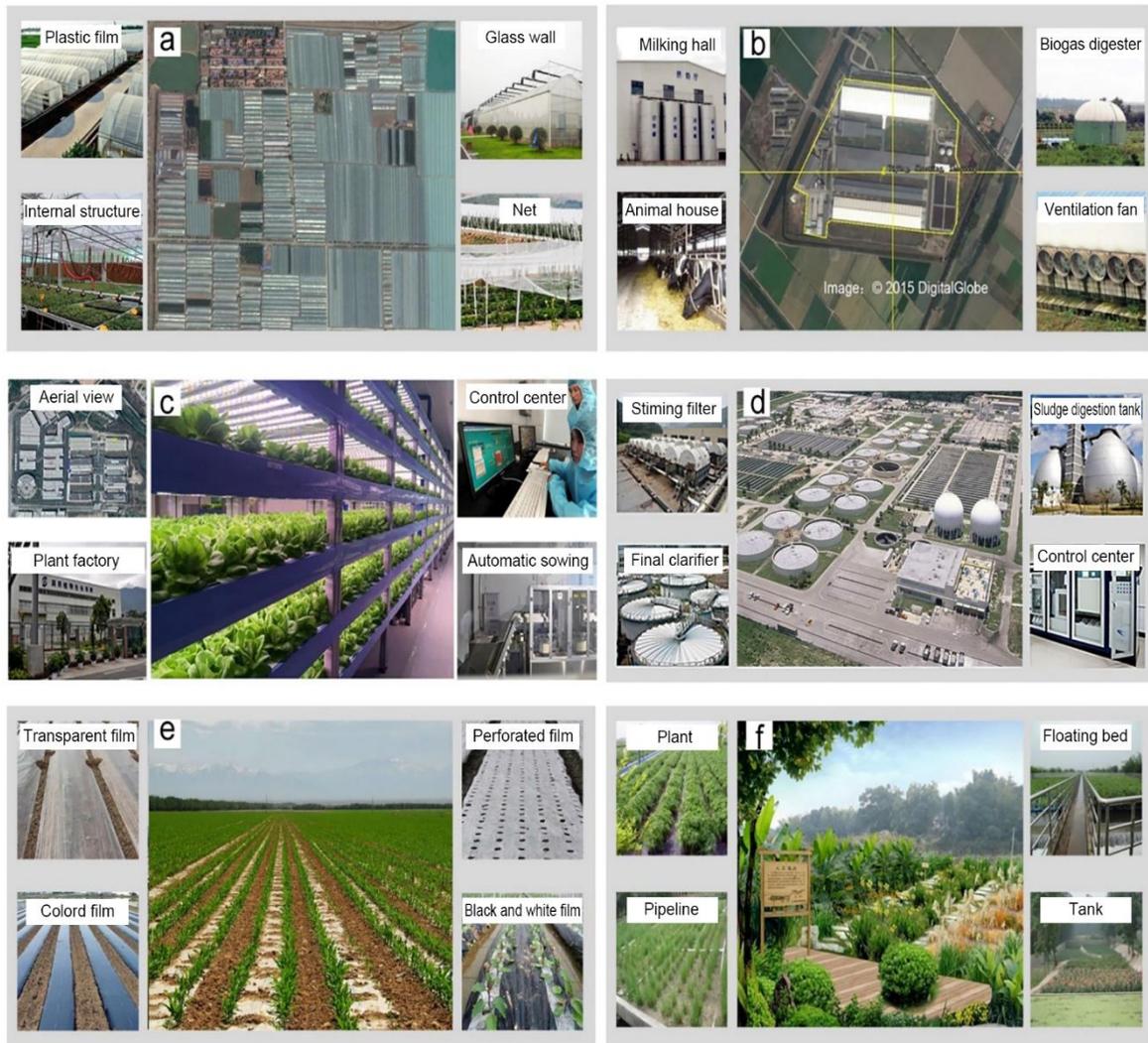

**Figure S3. Structures of enclosed ecosystems and other artificial ecosystems. a.** Greenhouses for vegetable, fruit or flower production; more than 90% of the greenhouse areas worldwide are covered by plastic films and the remainder are covered by glass or plastic net. **b.** Dairy feedlots, the cow concentrated areas for feeding and milk production. **c.** Vertical farm, a highly industrialized ecosystem that completely controls environmental conditions via multi-tiered cultivation. **d.** Wastewater treatment plants are composed of sediment tanks, bioreactor tanks (e.g., anaerobic and aerobic digestion tanks), stirring filters, sludge digestion tanks, final clarify tanks, as well as pumping stations and control rooms; many tanks have top covers and a bottom liner. **e.** Mulched farmlands. **f.** Constructed wetlands for wastewater treatment are fed with wastewater, have plants living on the surface and have water-proof bottom covers and side walls.

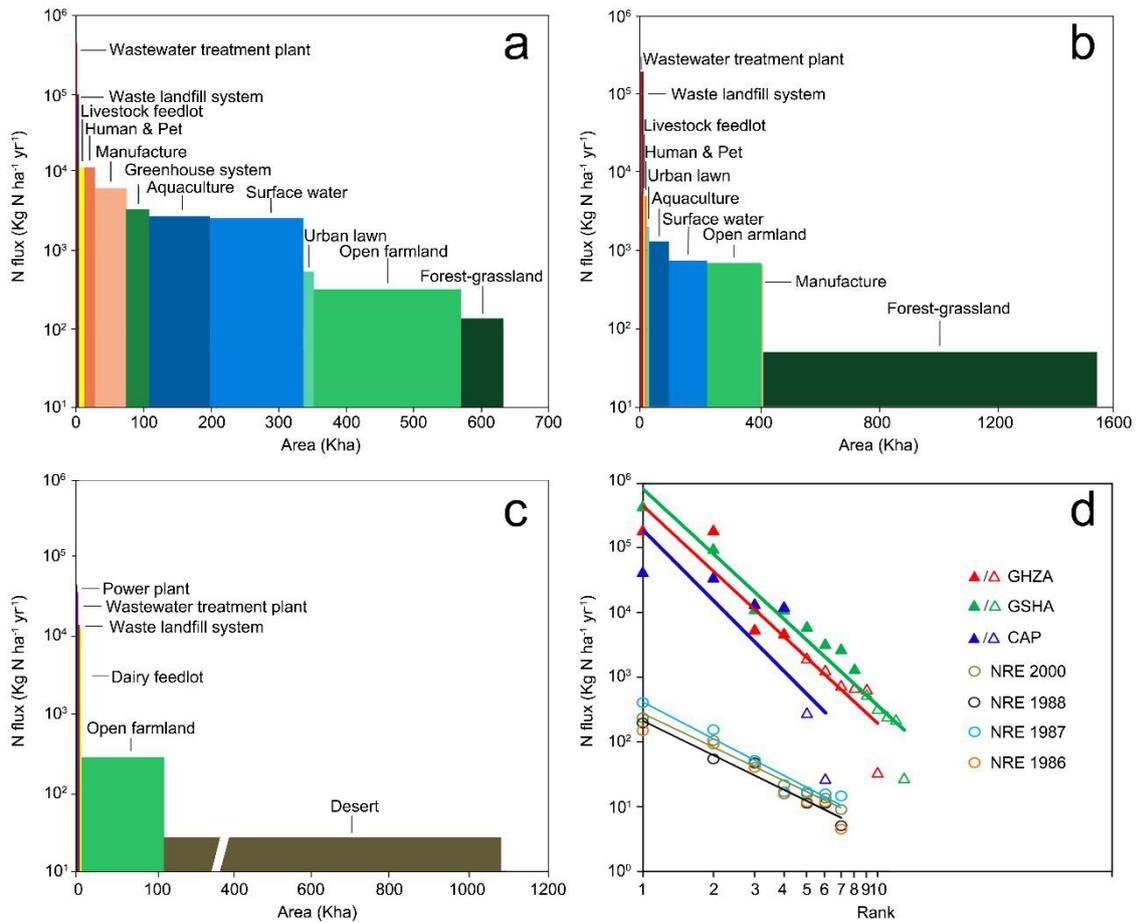

**Figure S4. Nitrogen fluxes in the case study cities and ecosystems. a-c.** Nitrogen flux-land occupation of the enclosed ecosystems and other ecosystems in Greater Shanghai Area (GSHA), Greater Hangzhou Area (GHZA) of China, and Central Arizona-Phoenix (CAP) of USA; the widths of the bars correspond to the land area of the ecosystems occupied and they are ranked by their nitrogen flux. **d.** Nitrogen flux distributions of metabolic nodes in the three cities (a-c) and a natural ecosystem, Neuse River Estuary (NRE) of USA; closed triangles are the enclosed ecosystems while open triangles are the open ecosystems in three case cities (a-c), and the open circles are the trophic grades in river ecosystem in different year.

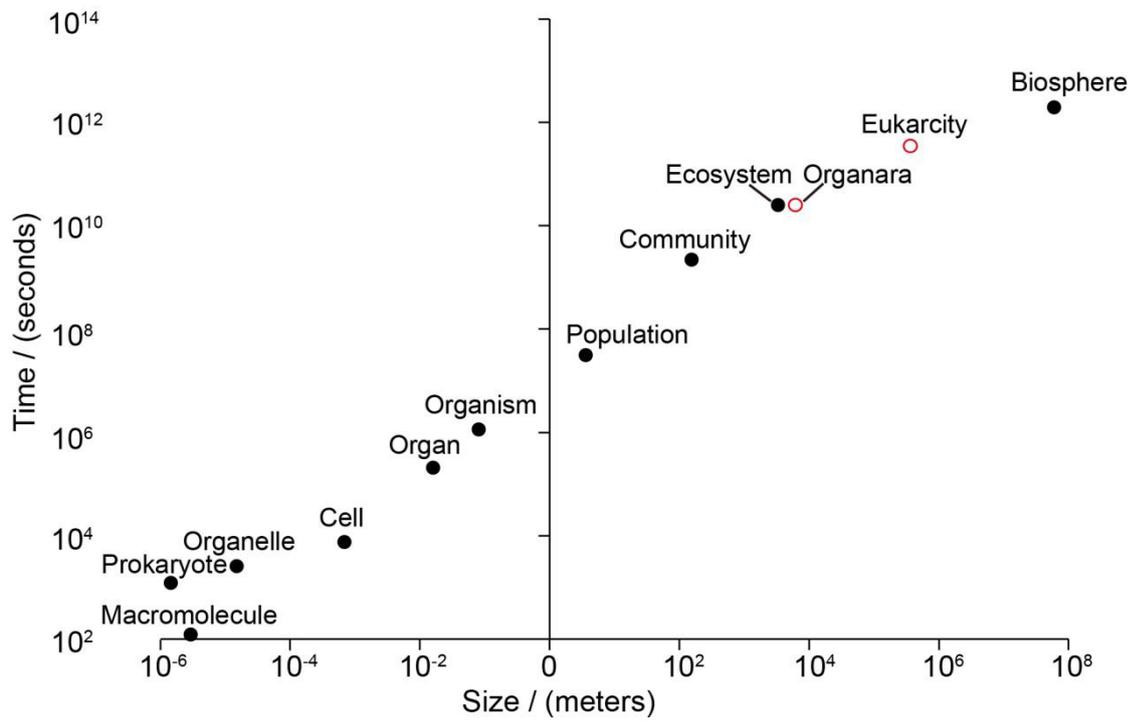

**Figure S5. Hierarchical living systems from macromolecule to biosphere.** Closed circles denote existed living systems, open circles denote the newly emerged eukarcity and organara. In X-axis, size is a space across the system of certain level; in Y-axis, time is action duration of entities including interaction, division, life span, generation, renewable, regeneration, changing and evolutionary time.

**Table S1. Examples of enclosed ecosystems and the functional groups**

| Category | System example | Functioning | Location in a city |
|---|---|---|---|
| **Biological producer** | | | |
| Primary | Greenhouses, vertical farms | Supply vegetables and fruits | Suburban and agricultural areas |
| Advanced | Industrial livestock feedlots | Supply meat, eggs, milk, aquatic products | Suburban |
| End processer | Restaurants, Breweries | Supply foods and drinks | Urban areas |
| **Non-biological producer** | | | |
| Primary | Mines, oil/gas fields, Reservoirs | Supply minerals, energy and water | Rural areas |
| Advanced | Power plants, manufacturing plants, waterworks, electrical appliance factory | Supply electricity, heat, material, food, supply products for production and life | Exurban areas |
| **Decomposer** | | | |
| Wastewater | Wastewater treatment plants | Purify wastewater | Exurban areas |
| Solid waste | Garbage landfills, incineration plants | Waste disposal; Reusing resources (material cycle) | Urban fringes |
| Waste gas | Gas purification units | Purify waste gas | |
| **Transporter** | | | |
| | Stations, crafts, ports | Transfer people, goods, energy | Inside / among cities |
| **Information delivery** | | | |
| | Telecommunication bureaus, post offices, network servers | Communication | Concentrated in urban areas |
| **Carrier** | | | |
| | Stores, warehouses, restaurants, gasoline stations | Link production and consumption | Concentrated in urban areas |
| **Regulator** | | | |
| | Governments, banks, hospitals | Manage and control sub-systems | Urban areas |
| **Information** | | | |

| | | |
|---|---|---|
| Junior and senior schools, universities, research institutes, statistics bureaus, museums | Save, copy, innovations | Concentrated in urban areas |

**Table S2. The 15 traits of living systems, city and city components for clustering analysis**

| Type | Trait | Organelle | Eukaryotic cell | Organ | Organism | City component* | City** | Assignment |
|---|---|---|---|---|---|---|---|---|
| Morphology | Shape[1, 2, 3] | 2 | 3 | 2 | 1 | 2 | 3 | Sphere (3), near sphere (2), not sphere (1) |
|  | Diversity | 2 | 3 | 1 | 7 | 3 | 2 | Values are the log number of 'species' in morphology; e.g. it is estimated that there are at least 10 million ($10^7$) species |
|  | With outer membrane | 1 | 2 | 1 | 2 | 1 | 0 | All types of systems have (2), only some systems have (1), without (0) |
| Structure | Fixed number of components in a system | 0 | 0 | 1 | 1 | 0 | 0 | Fixed (1), with difference among systems (0) |
|  | Number of components[4, 5] | 2 | 5 | 0 | 1 | 2 | 4 | Log number, i.e. there are ~ 100 types of organelles |
|  | With dynamics of component quantity | 1 | 2 | 0 | 0 | 1 | 2 | Individuals of the systems have competition and cooperation. Number of individuals of a system changes greatly (2), medium (1), and weak (0) |
|  | With control center | 1 | 2 | 0 | 2 | 1 | 2 | A system has nucleus (2), with control center but no nucleus (1), almost no control center (0) |
|  | With hotspots in matrix[6] | 2 | 3 | 0 | 0 | 1 | 3 | Difference between hotspot and matrix is very large (3), large (2), medium (1), and small (0) |
|  | Regeneration capacity | 0 | 0 | 1 | 0 | 0 | 1 | Some part of system can regenerate after removal (1) (for example lizard's tail), and cannot regenerate |
| | | | | | | | | (0) |
| Process | Internal steady state | 0 | 0 | 0 | 1 | 1 | 1 | With homeostasis mechanisms for temperature (1), without (0) |
| | Be controlled by high level of system | 1 | 0 | 1 | 0 | 1 | 0 | Be controlled (1), not (0); e.g. an organelle's metabolism is controlled by nucleus, while an organism's metabolism is not controlled by a population |
| | With life history | 0 | 1 | 0 | 1 | 1 | 1 | With procedural death (1), without (0) |
| | Metabolism rate in response to system size[7, 8] | 3/4 | 3/4 | 5/6 | 3/4 | 5/6 | 5/6 | The values are the power exponent ($\beta$) of the function $Y = aX^\beta$ |
| Function | Exchange materials and energy with other systems | 1 | 0 | 1 | 0 | 1 | 0 | Exchange frequently (1), almost no (0); they exchange with environment |
| | Degree of organization with information itself | 1 | 2 | 0 | 1 | 0 | 1 | Strong (2), medium (1), weak (0) |

[*] Equal to enclosed ecosystem; [**] Equal to urban-rural system

**Table S3. Characteristics and functions of the outer covers of enclosed ecosystems**

| System | Type | Materials | Position | Function |
|---|---|---|---|---|
| Greenhouse | Roof | Glass, plastic, straw, chemical fiber | Top cover | Maintain internal environment |
| | Wall | Glass, plastic, brick, earth | Side cover | Maintain internal environment |
| Dairy feedlot | Roof | Steel, tile, plastic, glass | Top cover | Maintain internal environment |
| | Wall | Brick, earth, wood, steel | Side wall | Maintain internal environment |
| | Fence | Steel, wood | Side wall | Protection |
| Wastewater treatment plant | Roof | Tile, cement, steel, plastic | Top cover | Maintain internal environment |
| | Wall | Cement, brick | Side wall | Maintain internal environment and waterproofing |
| | Bottom liner | Cement, steel | Bottom layer | Prevent wastewater leaching |
| Constructed wetland | Bottom liner | Geotextile, cement, earth | Bottom layer | Prevent wastewater leaching |
| | Wall | Cement, brick, geotextile | Side wall | Waterproofing and maintain internal environment |

**Data Sources for the Main Text Figures**

The data sources for Figure 1 to Figure 3 and Figure 5 in the main text.

**Figure 1**

B shows the sampling site for the six types of ecosystems used for case studies showed in A and C. The data are collected from the literature that covered China, USA, EU, and so on[9-13]. The 15 traits of six levels of living systems are list in D, while data are listed in table S3. Data source for E is based on table S3.

**Figure 2**

A is drawn by using Photoshop CS6, and the data for B are extracted from research[14] by using GetData Graph Digitizer, C is extracted from research[15]. Data sources for D to F are based on previous research[16] and our investigations. G is compiled according to related literature[17-21]. The number of greenhouses and wastewater treatment plants and the human population for H to I were obtained from a statistical yearbook[22] and government statistic websites[23, 24].

**Figure 3, B to E**

Data sources for the monthly variations of internal and external[25] air temperatures of enclosed ecosystems; greenhouses in De Blit, the Netherlands[26], a vertical farm (crop factory) in Anxi county, south China (this study), dairy feedlots in Ottawa, Canada[27], and dairy feedlots in Shanghai, Southeast China[28].

**Figure 5**

D and F: The per capita GDP and population of Lake Taihu basin are from[29]. E: The cumulative number of WTPs in cities around Lake Taihu[30]. G to I: Data sources for these figures include the dynamics of open farmlands, greenhouses, dairy feedlots, forests, wetlands, GDP and sustainable grassland area in China from 1980s to 2010s[9, 31-33].

**Supplemental Reference**